\documentclass[12pt]{iopart}
\usepackage[english]{babel}
\usepackage[utf8x]{inputenc}
\usepackage[dvipdf]{graphicx}

\begin{document}

\title{Cutting and controlled modification of graphene with ion beams}
\author{O. Lehtinen,$^{1}$ J. Kotakoski,$^{1}$
  A.~V. Krasheninnikov$^{1,2}$ and J. Keinonen$^{1}$}

\address{$^{1}$ Department of Physics, P.O. Box 43, FI-00014
  University of Helsinki, Finland\\ $^{2}$ Department of Applied
  Physics, P.O. Box 1100, FI-00076 Aalto University, Finland}

 \date{\today}

  \begin{abstract}
    Using atomistic computer simulations, we study how ion irradiation
    can be used to alter the morphology of a graphene monolayer by
    introducing defects of specific type, and to cut graphene sheets. 
    Based on the results of our analytical potential molecular dynamics 
    simulations, a
    kinetic Monte Carlo code is developed for modelling morphological
    changes in a graphene monolayer under irradiation at macroscopic
    time scales. Impacts of He, Ne, Ar, Kr, Xe and Ga ions with kinetic energies ranging from tens of eV to 10~MeV and angles of incidence between 0$^\circ$ and 88$^\circ$ are studied. Our results provide microscopic
    insights into the response of graphene to ion irradiation and
    can directly be used for the optimization
    of graphene cutting and patterning with focused ion beams.
  \end{abstract}



\maketitle

\section{Introduction}

Recent reports on large-scale
production of graphene~\cite{choucair2009,xiaohong2010,bae2010}, including growth of
centimeter-size sheets on copper surfaces
\cite{XuesongLi06052009}, have brought closer the utilization of graphene's excellent electronic properties~\cite{Geim07rev} in future electronic devices.
In particular, quantum dots~\cite{Ponomarenko04182008},
nanoribbons~\cite{XiaolinLi02292008}, and antidot
lattices~\cite{bai2010,pedersen2008}, which provide electron confinement within the
graphene plane, have received considerable attention. Production of
such structures is based on graphene patterning by various techniques
such as electron beam lithography combined with reactive ion
etching~\cite{Ponomarenko04182008,PhysRevLett.98.206805,stampfer:012102},
chemical methods including unfolding of carbon
nanotubes~\cite{kosynkin2009,jiao2009} and graphene cutting with
a focused electron beam~\cite{fischbein:113107}.

Since focused ion beams (FIB) are already routinely employed in
today's semiconductor industry, this method could become an alternative
approach. Indeed, cutting and patterning graphene using a FIB with a
high spatial resolution was recently
demonstrated~\cite{lemme09,lemme09b,pickard2010}. The method is based on using a
30~keV He ion beam in a helium ion microscope to directly sputter
carbon atoms from predetermined areas of graphene sheets, either
suspended or deposited on a substrate. Etched gaps down to 10~nm have been
reported, with sharp edges and little evident damage or doping to the
sample. Concurrently with the experiments on He beam--based patterning
of graphene, the possibility of using Ar ion irradiation has also been
studied~\cite{PrivateCommJani}.

Efficient use of ion beams and optimization of the graphene cutting process
require detailed microscopic knowledge of damage production
mechanisms and types of defects created by the energetic ions in the 
sample. In order to get insight into the cutting process, the interaction of He
ions with the target was modeled in Ref.~\cite{lemme09b} by a
semi-empirical method~\cite{ZBL} based on the binary-collision approximation,
combined with statistical algorithms to calculate how a moving ion
transfers its energy to the target atoms. This approach
implemented in the TRIM software package \cite{SRIM2003} gives
reasonable results for bulk materials. However, as has been pointed
out \cite{Kra07nm,Kra10JAP}, the theory of irradiation effects
developed for bulk materials does not always work at the nanoscale. In particular,
it was recently demonstrated \cite{ossi2010} that this approach cannot
be applied to graphene, as the sample is treated as an amorphous
matrix with a homogeneous mass density while the explicit account for
the atomic structure is required for atomically thin targets.

In this study, we use analytical potential (AP) molecular dynamics (MD)
simulations, a much more accurate method than the one implemented in
the TRIM code, to study defect production in graphene under ion
irradiation. Our ultimate goal is to provide the means for determining
optimum parameters, such as ion mass, energy and angle of incidence
for graphene cutting, which would enable the production of smooth edges with
the minimum number of defects
at faster cutting rates. We simulated more than two million impacts of
energetic ions onto suspended graphene and gathered statistics on
types and abundances of defects for He, Ne, Ar, Kr, Xe and Ga ions with
energies ranging from tens of eV to 10 MeV. The role of the angle
of incidence of the ions was studied in detail. To establish a direct link to the
experiments, we
further developed a kinetic Monte Carlo (kMC) code~\cite{ikmc}, which
utilizes the statistics from the MD simulations for predicting the
evolution of graphene under ion irradiation at macroscopic timescales. 
This allows modelling the behavior of irradiated graphene under realistic experimental conditions in, {\it e.g.}, FIB systems.

\section{Methods}

In our MD simulations (a general introduction to this method can be found in Ref.~\cite{allentildeslay}) the carbon-carbon interaction in graphene
was modelled using the second-generation reactive empirical
bond-order Brenner potential~\cite{Bre02}. The
bond conjugation term, which is not expected
to be important in irradiation processes~\cite{Kra01prb}, was omitted.
This approach has previously been successfully 
used in modelling the irradiation response of graphene and
other carbon
nanostructures~\cite{ossi2010,Kotakoski05prb,Pregler06prb,Tolvanen07,xu:043501}.
The
interaction between energetic noble gas ions (He, Ne, Ar, Kr
and Xe) and  target carbon atoms was simulated using the
Ziegler-Biersack-Littmark universal repulsive
potential~\cite{ZBL}. In addition to the noble gas ions, irradiation
with Ga ions was simulated using the same universal repulsive
potential, as gallium is a typical ion species used in FIB
systems. Chemical effects are expected to be important only at low ion energies ($<250$~eV) and they should not play any role at typical operating energies ($\sim$30~keV) of FIB systems, especially in such a thin target,
so that the use of
the purely repulsive potential is justified also for the Ga--C interaction. 

For a direct analogy with the experiments, we use the term "ion" throughout
this article, although the charge of the incoming ion is not
explicitly considered, as this is beyond the AP-MD method, and more
importantly, the effects of low charge states are negligible~\cite{Kra07prl}. The AP-MD approach is computationally efficient enough for running the
massive number of simulations (more than two million runs in total)
required to get statistically meaningful results in the large
parameter space explored.

It is well known that electronic stopping is the main mechanism of
energy transfer from an ion to any solid target at high ion energies
\cite{ZBL}.  At the same time,
experiments~\cite{PhysRevB.64.184115,Caron200636} indicate that energy
deposited into electronic degrees of freedom of graphite gives rise to
defects only if electronic stopping power exceeds a critical value of
7~keV/nm. Taking into account the excellent charge and heat
conductance of graphene, similar behaviour can be expected for this
material~\cite{Kra07prl}. As the typical electronic stopping value for
all the ion/energy combinations used in our study is less than
0.7~keV/nm (we calculated electronic stopping power using the approach of Ziegler, Biersack, and Littmark~\cite{ZBL}), and even the highest values for high-energy Xe 
($\sim 4.5$~keV/nm)
are well below the critical one, electronic stopping effects were not
taken into account.

The simulated graphene target consisted of 800 carbon atoms. Since thermal excitations due to temperature-related atomic motion do not play a significant role in the momentum transfer between the impinging ion and target atoms (typical phonon energies are within tens of meV per atom), the initial target temperature was chosen to be 0~K.
After ion impacts the system was cooled down using the Berendsen
thermostat~\cite{Ber84} at the edges, with $\Delta \tau=10$~fs as the
time constant. Such a setup was used to model the dissipation of heat generated by
the ion impact as would happen if the target was part of a larger
graphene sheet. Additional simulations with the thermostat turned off
showed, however, that the results are not sensitive to its
parametrization, as the damage creation processes typically take place
before the heat wave generated by the ion impact even reaches the
system edges, barring a few cases in the high ion energy and high
angle of incidence regime. Adaptive time step MD code {\sc parcas}~\cite{parcas}
was used for the simulations. Within the code, the simulation time step
is dynamically adjusted based on the fastest moving particles in the
system. This resulted in time steps ranging from the attosecond scale
to close to one femtosecond.

The system was allowed to relax for one picosecond after the ion
impact. During this time the system had typically reached a local energy
minimum. However, any reasonable simulation time is too short for the
system to always find the most stable local configuration. To address
this, the resulting structures were annealed for another picosecond at
1500~K and eventually cooled down to 0~K to facilitate further
relaxation of the system and removal of at least the most unstable
atomic configurations. Over macroscopic time scales at experimentally relevant temperatures (room temperature and above) the created defects are expected to reconstruct into patches of non-hexagonal carbon rings similar to those presented in Ref.~\cite{kotakoski2011}. Although the AP-MD method cannot capture the reconstruction processes, the sizes of these patches are defined by the initial damage which is statistically predicted by our results.

\begin{figure}
  \includegraphics[width=.48\textwidth]{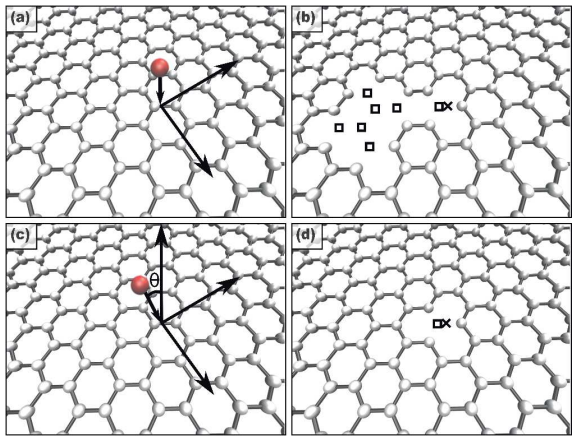}
  \caption{Ball-and-stick presentation of the
    simulation setup. (a) Ion impact in direction perpendicular to the
    graphene plane. (b) Structure of the graphene target after the impact (complex vacancy), 
    where the damage is caused by an in-plane collision cascade. (c) Ion
    impact at an angle $\theta=30^\circ$ toward exactly the same spot
    as in (a). (d) Structure after the impact (single vacancy), where the C atom is recoiled out of the plane, thus creating no further damage. The crosses mark the
    ion impact points and the squares the original locations of
    sputtered atoms. Both impacts were carried out with 2~MeV Ar$^+$
    ions.}
  \label{schema}
\end{figure}

Impact points for the ions were chosen randomly within the minimum
irreducible area of the graphene lattice. The direction of the
incoming ion was determined by randomly selecting an inclination angle
$\theta \in [0,88^\circ]$ and azimuthal angle $\varphi \in
[0,360^\circ[$ independently so that for each $\theta$ an average
result could be generated over $\varphi$. A schematic illustration of
the simulation setup is presented in Fig.~\ref{schema} along with an
example of the effect of changing the $\theta$ angle.

A link between the theoretical
data on damage production and experiments can be provided within the framework of
the kMC method. We have previously implemented such a model for
comparing theoretical results to experiments in the cases of electron
irradiation of carbon nanotubes~\cite{Gan07} and
hexagonal boron nitride monolayers~\cite{Kotakoski10BN}.
Here, we developed a kMC model describing the response of graphene to ion
irradiation by using the results
from our AP-MD simulations as the source of damage event rates. 
Within this model, the
impact event rate is derived from  ion beam current, and the rates
of producing any type of defects are derived by multiplying the ion
impact rate by defect creation probability for the selected combination
of ion species, irradiation energy and angle of incidence. We
employed data on the defect size (area initially covered by perfect hexagonal carbon rings transformed into other polygons, termed "amorphized area" within our kMC model)
and the number of sputtered carbon atoms for each parameter
combination. Because the MD simulations were always carried out on a
pristine target, one should be careful when interpreting
the kMC results when the defects start to overlap. However, as it is
easier to displace under-coordinated atoms, our kMC results can be
used to estimate the lower limit for the number of sputtered atoms even
for overlapping defects.

\section{Results and Discussion}

\subsection{Dynamical simulations of defect production}

\begin{figure*}
  \includegraphics[width=.98\textwidth]{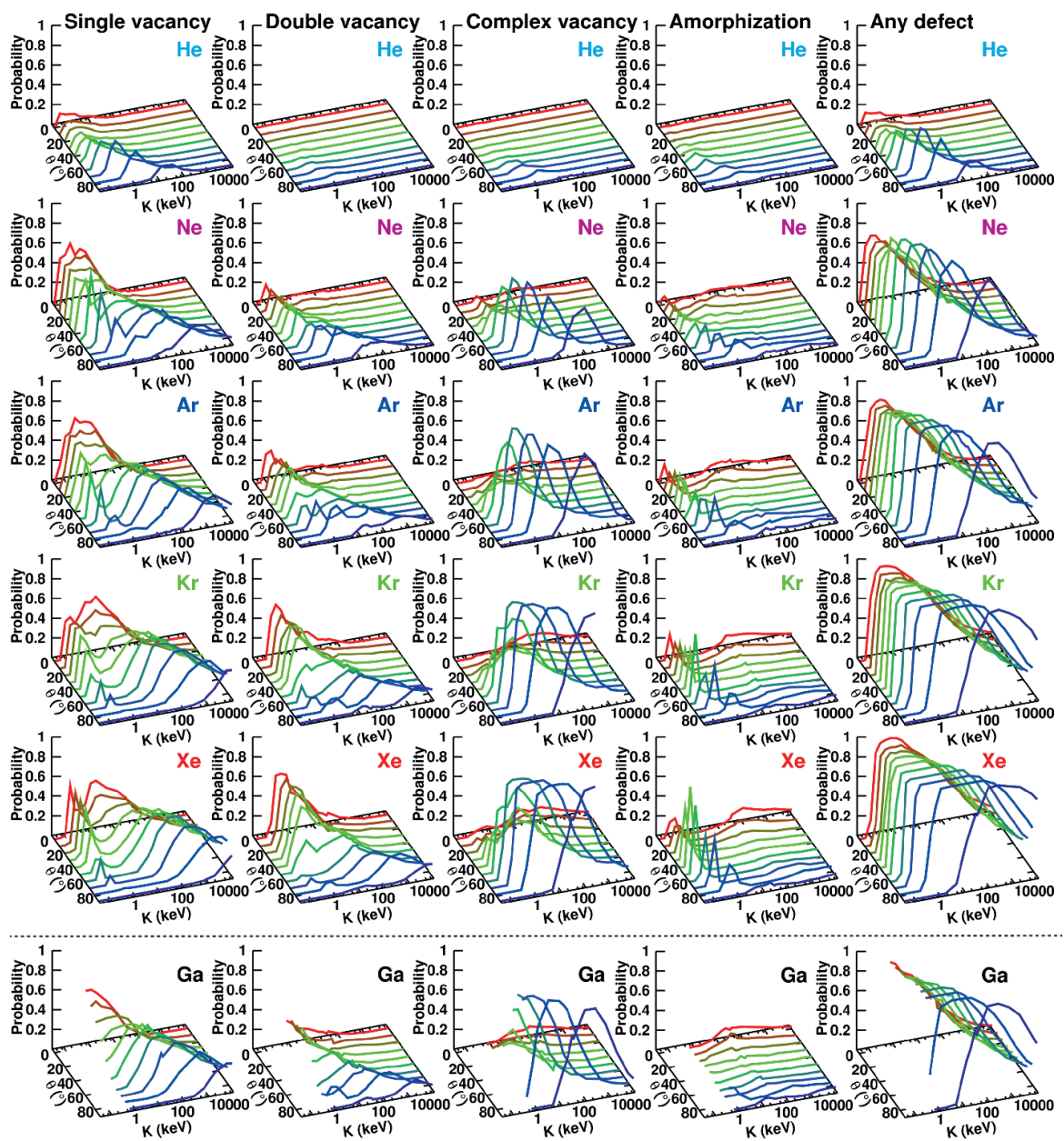}
  \caption{Probabilities of producing single
    vacancies, double vacancies, complex vacancies, local amorphizations and any defect (any modification to the pristine structure) in graphene under ion irradiation as functions of angle of incidence $\theta$ and ion
    energy $K$.  Data for Ga is presented only for energies $K\geq$0.5~keV. Data for lower energies is excluded as chemical effects were not accounted for in the Ga--C interactions.}
  \label{defprobs}
\end{figure*}

When analyzing the results of the MD simulations, we categorized the
defect structures produced in graphene into single vacancies (one missing atom
from otherwise intact structure), double vacancies (two missing
atoms and no further damage), complex vacancies (defect structures
with missing atoms other than single and double vacancies) and amorphous regions (defect
structures with no missing atoms, the simplest example of such a structure being the Stone-Wales defect~\cite{stone1986}). Experimental examples of these structures are presented in a recent review article~\cite{banhart2011}, except for complex vacancies, examples of which are presented in Ref.~\cite{kotakoski2011}.
We calculated the probabilities of
producing each of these defect types for all ion species and the whole
range of energies (35~eV -- 10 MeV) and 
angles of incidence ($\theta \in [0,88^\circ]$) considered, as presented in
Fig.~\ref{defprobs}. Also the probabilities of creating any defect was calculated.

As is evident from Fig.~\ref{defprobs}, varying any of the parameters
has a drastic effect on the produced defect types and their
abundances. At energies in the keV range, single vacancies are the typical defects
when the angle of incidence is perpendicular to the graphene sheet.
Upon tilting the ion beam, first double vacancies become the dominant defect type after
which complex vacancies become the most common type of defect. This
can be attributed to the increased projected density of graphene, when
viewed from a grazing angle.  The maxima in single vacancy production are shifted
towards higher ion energies as the angle of incidence is increased,
which can be useful when ion irradiation with a fixed
energy is used for cutting graphene. 
The location of the maxima in complex vacancy production first
moves toward lower energies with increasing angle of incidence after
which it again moves to higher energies. Therefore, if graphene is to be used as
a membrane in external beam experiments as suggested
in Ref.~\cite{ossi2010}, the resilience of the membrane can be improved by
slightly tilting it. For example, in the case of 2~MeV Ne ions, tilting the graphene sheet by $\sim 18^\circ$ will result in a twofold decrease in the sputtering rate, as will be discussed in more detail later in this article.

A range of parameters where no defects are created corresponds to the large angle,
low energy region of the graphs in Fig.~\ref{defprobs}. Low defect production
has the same origin as under ion channeling conditions in crystalline solids. When coming in at a
grazing angle, the ion interacts with a long row of atoms, and if the
energy is low enough, depending on the angle of incidence, none of the target atoms
receives enough momentum to be displaced. Instead, the ion is reflected away
from the graphene sheet. However, if ion energy is increased, the ion
will penetrate the sheet, and it typically creates significant damage along the
way, as can be seen from the steeply rising complex vacancy
probabilities in the low energy--large angle regimes in
Fig.~\ref{defprobs}. Amorphization events are observed mostly at low ion energies ($< \sim
$ 1~keV), where the ions have barely enough kinetic energy to displace a target atom, but the displaced atom remains bonded to the
sheet as an adatom or a part of a local amorphization.

\subsection{Statistical model for continuous defect production}

General trends and dominant defect types for specific irradiation
conditions can be directly obtained from Fig.~\ref{defprobs}.
However, making quantitative predictions on the evolution of a
graphene target under ion irradiation is not straightforward, as both
variations in the defect sizes and the number of sputtered
atoms must be taken into account. Also, the probability distribution
for creating each defect is not Gaussian, which means that mean size
and standard deviation are not sufficient for describing the damage
caused to graphene. For this reason, we developed a
kMC code, which directly uses the statistics on the sizes
of the defects and number of sputtered atoms, extracted from the MD
simulations. This code
can be used for predicting the evolution of a
graphene sheet under ion irradiation at macroscopic time scales. The code has a web-based interface~\cite{ikmc} and is available for public use. Both
a visual representation of the sample and quantitative estimates of
the total amorphized area and the number of sputtered atoms can be
obtained using the program. As the original data is collected for
individual ion impacts on pristine graphene, we stress that the code
underestimates the produced damage at high ion doses due to a drop in
the displacement threshold energy for under-coordinated atoms~\cite{Kra10JAP}.

\begin{figure}
  \includegraphics[width=.48\textwidth]{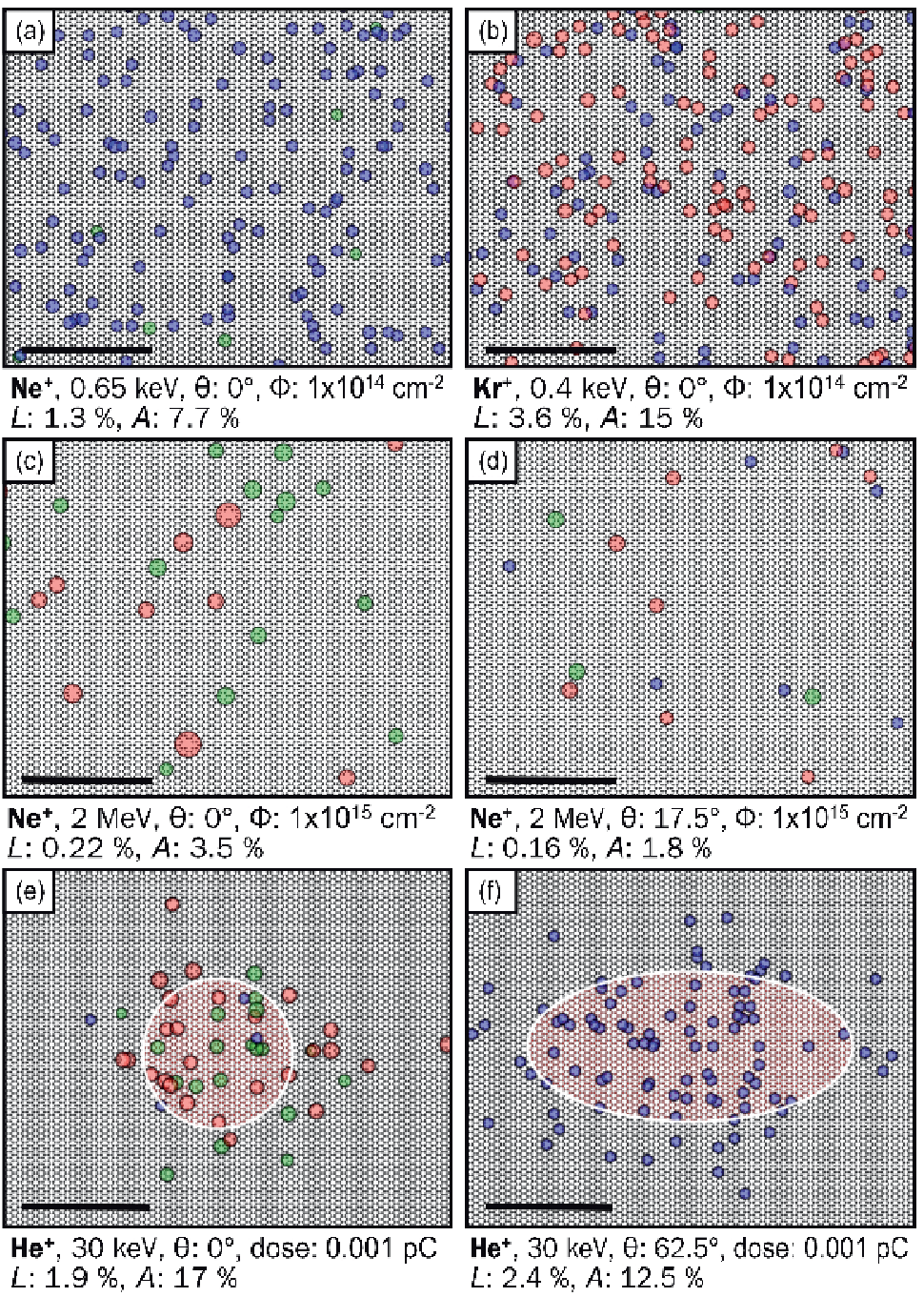}
  \caption{Example results of kinetic Monte Carlo simulations of
    graphene evolution under ion irradiation. Blue circles mark
    single vacancies, whereas the red ones stand for other
    vacancy-type defects, and the green ones for amorphized areas with
    no missing atoms. Percentages of lost carbon atoms ($L$) and amorphized
    area ($A$) are given for each case. Note that vacancy-type defects also
    contribute to the percentage of amorphized area. All the numerical 
 values were averaged over 25 simulations. The length of the scale bars is 5 nm in all the panels. The pink areas in panels (e) and (f) indicate the spot area of a focused ion beam limited at the full width at half maximum of the spot intensity. $\Phi$ is irradiation fluence as in ion impacts per unit area and dose stands for total accumulated charge of the impinging ions.}
  \label{kmc_a}
\end{figure}

\begin{figure}
  \includegraphics[width=.48\textwidth]{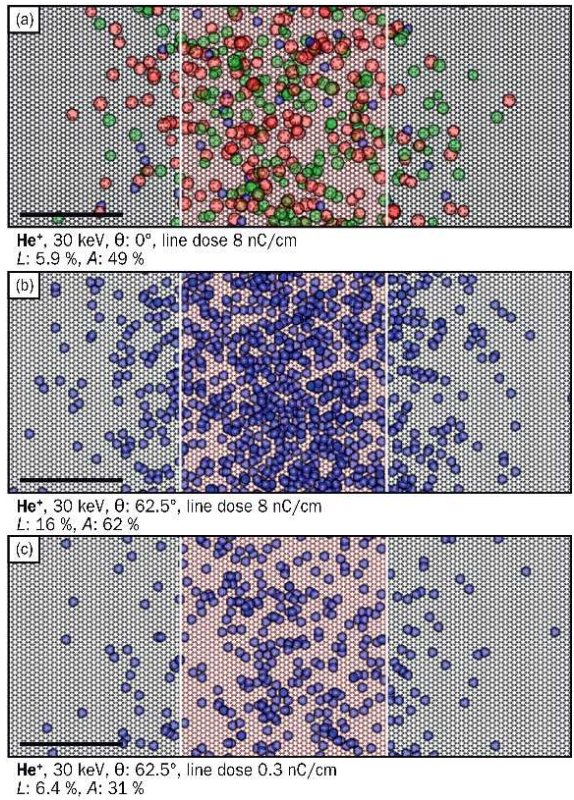}
  \caption{Results of kMC simulations of graphene evolution under ion irradiation 
 with parametrization similar to what was used in experiments in Ref.~\cite{lemme09} (panel (a))
and with optimized irradiation angle with regard to cutting efficiency (panels (b) and (c)). Blue circles mark
    single vacancies,  the red ones stand for other
    vacancy-type defects, and the green ones for amorphized areas with
    no missing atoms. The pink area indicates the area swept by the ion beam limited at the full width at half maximum of the beam spot. Percentages of lost carbon atoms ($L$) and amorphized
    area ($A$) are given for each case.  The scale bars are 5 nm. It is evident that by tilting the ion beam, the sputtering rate is increased significantly and the creation of local amorphizations is avoided. Line dose stands for total accumulated charge through charged ion impacts per unit length.}
  \label{kmc_b}
\end{figure}

To demonstrate the potential of the kMC code, in Figs. \ref{kmc_a} and  \ref{kmc_b} 
we present example
cases simulated with the code. 
If graphene is uniformly irradiated with 650 eV Ne ions at normal
direction to the sheet, nearly exclusive production of single vacancies is observed
(Fig.~\ref{kmc_a}a). Changing the ion to 400 eV Kr will result in
uniform distribution of single and double vacancies with roughly a 1/1 ratio
(Fig.~\ref{kmc_a}(b)). For high energy ions with incident direction
normal to the sheet the probability of creating any defect is greatly
reduced, as is evident from Fig.~\ref{kmc_a}(c) for the case of 2~MeV Ne: 
Although the fluence is increased by one order of
magnitude as compared to the previous examples, the total number of
defects is lower. The individual defects are much larger, however,
which can be attributed to the fact that during the interaction of a
target atom and the fast ion the target atom is practically immobile,
resulting in symmetrical momentum transfer in the direction of the ion trajectory. This leads to recoiled atoms moving very close to a direction perpendicular to the ion trajectory. In the case of ion trajectory perpendicular to the graphene sheet,
this leads to in-plane 
collision cascades and large defect structures. This effect is also
demonstrated in Fig.~\ref{schema}. If the direction of the ions is
tilted away from the normal of the graphene plane, the typical direction of the
recoiled atoms is correspondingly tilted out of the graphene plane,
which results in smaller defect structures due to the absence of
collision cascades, as can be seen in Fig.~\ref{kmc_a}(d). Thus the
resilience of graphene under high energy ion irradiation can be
improved by tilting it, as was suggested above.

\subsection{Optimizing ion beam cutting of graphene}

As experimentally demonstrated by Lemme {\it et
  al.}~\cite{lemme09} and Pickard {\it et al.}~\cite{pickard2010}, a
focused He ion beam can be used to cut graphene with a very high 
precision. The used acceleration voltage was 30~kV and the cutting was
performed with the ion beam pointed in the normal direction of the
sample surface. The effects of such irradiation on graphene are shown in
Fig.~\ref{kmc_a}(e) assuming a focused spot with a Gaussian distribution with a
full width at half maximum (FWHM) of 6 nm.  
Approximately half of the damage events lead to sputtering of
target atoms and half of the events lead to amorphizations. If clean
cutting is to be achieved, the amorphization events are not
desirable. The situation can be improved by tilting the beam by
approximately 60$^\circ$ (Fig.~\ref{kmc_a}(f)). As the projected
atomic density of the target is now increased, the defect creation
probability is correspondingly higher. The percentage of sputtered
atoms inside the spot area (illustrated with a white line along the
FWHM edge) is increased with the same total dose even though the spot
area is larger. Also, as demonstrated previously, tilting the beam at high ion energies 
leads to decreased probability of creating in-plane collision cascades and local amorphizations, in favor of single vacancies. This leads ultimately to cleaner edges in cuts made with a FIB.

To further illustrate what can be achieved with parameter optimization when
cutting graphene with a FIB, three examples of the use of the
kMC code in linear scan mode are presented in Fig.~\ref{kmc_b}. In these
examples, the beam spot is moved in the vertical
direction. The spot width is 10 nm (FWHM) and the ion beam direction is tilted
towards the direction of the spot movement. 
The ion energy and species are
similar to what was used in Ref.~\cite{lemme09}. Line dose of 8
nC/cm was reported to not be adequate for making a cut in
graphene. This observation is supported by our simulations
(Fig.~\ref{kmc_b}(a)): Only $\sim$6~\% of the carbon atoms within the
cut area are sputtered, although much of the area is amorphized. However, when
the ion beam is tilted by 62.5$^\circ$ there is almost a threefold
increase in the number of sputtered atoms (Fig.~\ref{kmc_b}(b)). To
compare the effectiveness of graphene cutting at different
angles of incidence, a third simulation was conducted, where the line
dose was decreased so that the number of sputtered atoms is approximately the same
as in the first case. Comparing the results in Figs.~\ref{kmc_b}(a)
and~\ref{kmc_b}(c) shows that tilting the He beam by $\sim$60$^\circ$
gives a threefold increase in the cutting efficiency of the beam,
while the amorphized area outside the cut is decreased considerably.

\section{Conclusions}

To conclude, we studied the effects of ion irradiation on graphene
using atomistic computer simulations. The role of the angle of ion
incidence in a wide range of energies was investigated, and the types
and concentrations of defects were identified for various ions. The
dramatic effect of the angle of incidence on defect production
demonstrates the fundamental difference of strictly two-dimensional
graphene from traditional bulk targets. The peculiarities of
graphene's response to ion irradiation can be used to gain detailed
control over produced defect types and their abundances. The presented publicly available computer code~\cite{ikmc} enables one to make quantitative predictions of defect production in
graphene under energetic ion bombardment. This information is needed
in order to controllably create specific types of defects, or when graphene is to be nanomachined by a FIB. To illustrate the latter, we showed an example
where $\sim 60^\circ$ tilting of the sample gave a threefold increase in
sputtering and reduced amorphized areas in the sample. Further on,
with respect to the possible use of graphene
membranes in ion-beam analysis~\cite{ossi2010}, we showed that the resilience of the
membrane can be improved for high energy ions by choosing an
optimum angle with regard to the direction of the ion beam.
Although we looked specifically at noble gas irradiation, our results can also 
provide insights into the response of graphene to irradiation by other species, e.g., nitrogen, as in very recent experiments on irradiation mediated nitrogen doping of graphene~\cite{guo2010}.

\section{Acknowledgments}
We thank the Finnish IT Center for Science for generous grants of
computer time. This work was supported by the Academy of Finland
through several projects and the Centre of Excellence programme.

\end{document}